\documentclass[journal=jpccck,manuscript=article]{achemso}
\usepackage{times}
\usepackage{epsfig}
\usepackage{graphicx}
\usepackage{psfrag}
\usepackage{ae}
\usepackage{amsmath,amssymb}
\usepackage[usenames]{color}
\usepackage{float}

\title{ Flexible Bond and Angle, FBA/$\epsilon $ model of water}

\author{ Ra\'ul Fuentes-Azcatl} 
\email{razcatl@xanum.uam.mx}
\affiliation{Instituto 
de F\'{\i}sica, Universidade Federal
do Rio Grande do Sul, Caixa Postal 15051, CEP 91501-970, 
Porto Alegre, RS, Brazil}
\author{Marcia C. Barbosa} 
\email{marcia.barbosa@ufrgs.br}
\affiliation{Instituto 
de F\'{\i}sica, Universidade Federal
do Rio Grande do Sul, Caixa Postal 15051, CEP 91501-970, 
Porto Alegre, RS, Brazil}
\begin{document}
\date{}
\begin{abstract}
We propose a new flexible force field for water. The model in 
addition to the Lennard-Jones and electrostatic parameters, includes
 the flexibility of the  OH bonds and  angles. The
 parameters are selected to give the 
experimental values of the density and dielectric constant
of water at at  $1$ bar at 240K
and  the dipole moment of minimum density. The FBA/$\epsilon $
reproduces the experimental values of structural, thermodynamic and the phase 
behavior of water in a wide range of temperatures with better accuracy than atomistic 
and other flexible models. We expect that this new
approach  would 
be suitable for studying water solutions.
\end{abstract}
\section{Introduction}
Water is ubiquitous in nature and strongly affects other materials
 when is in solution. This 
simple molecule, two hydrogen atoms linked to the oxygen by a covalent bond.
In gas phase  the  HOH angle is $104,474^o$ and the distance between oxygen
 and each hydrogen in  $0.095718\;nm$~\cite{Hasted}. This angular
 structure is not
 fixed. In the liquid phase
at  $298\;K $ and $1\;bar$ it reaches the 
angle of 106 degrees~\cite{Ichikawa}. Since the electrons in the 
covalent $HO$ bond are strongly attracted to the oxygen, water is 
polarized with the region of the oxygen negative and the region of
the hydrogen positive. Consequently
 the oxygen of one water attracts the hydrogen of the other
water molecule forming the hydrogen bonding. Water
due to the  HOH angular structure
can form up to four hydrogen bonds what  leads to the
tetrahedral structure which can aggregate in octamers.  Then a small angular
difference in the  HOH covalent bonds can be relevant to the
cluster structure of water. 

In order to give a description of thermodynamic
and dynamic properties of water, a number of non-polarizable models have been
developed. The idea  is to 
adjust the interaction potential  between the molecules so
the simulations reproduce the experimental value of a
a property such as the density
at a certain temperature and pressure. This process  has generated
rigid models, manageable computationally,  which give 
 accurate  values for a wide range of thermodynamic
and dynamic functions. However, since the potentials
are fitted at a  specific pressure and temperature, they are not able
 to cover different thermodynamic phases. 

This rigid models are unable
to capture  the changes in the water polarization  due to variations in 
temperature and pressure. This  becomes particularly problematic
when water is mixed with ionic or hydrophilic solutes.
For instance, the rigid models do not reproduce the
increase of the excess of specific heat when  alcohol ~\cite{Fu17}
is added to water
and they are unable to explain
 the enhancement of the  self-diffusion of water in the presence of certain
electrolytes ~\cite{Ki12}.
From the desire to produce an
atomistic model which accounts for changes
in the  HOH water angle, without the computational costs of the
ab initio simulations, a number of flexible models were introduced.
Some of them were based in original non-polarizable models
such as the SPC/E~\cite{Wu} and the TIP4P/2005~\cite{vegaf}
with the addition of more degrees of freedom.
 With this new approach, accurate water
 transport and other thermophysical properties
were obtained~\cite{Barrat, Smith, vegaf, aljf, Wu},
however, these models fail in reproduce 
other  thermodynamic
and dynamic properties~\cite{aljf, Wu}.
For instance, they do not give good estimates of the 
charge distribution~\cite{Warshel, Schmitt, Schmitt2,Wu,vega11}.
Therefore, a good  flexible model 
is still missing.

In this work we address this issue by following 
the same strategy adopted by Wu et al.~\cite{Wu} and
Gonzalez and Abascal~\cite{vegaf} by starting 
with a rigid non-polarizable model,the SPC/$\epsilon$~\cite{spce}  and 
introducing
flexible bonds and angles.  The idea is to combine the parametrization method
employed for rigid model with the flexibility which adds additional 
degrees of 
freedom for the parametrization. We selected the rigid 
SPC/$\epsilon$~\cite{spce}  as an starting point 
due to its simplicity and because 
the rigid model already give some thermodynamic~\cite{spce}
and dynamic properties~\cite{Az16} close to the 
experimental results at room temperature and pressures.

 The paper is organized as follows: Section 2 gives the force 
field of FBA/$\epsilon $ model of water, followed
 by the details of the simulation and pursuit the parameters, Section 3 
gives the properties of 
FBA/$\epsilon $ compared with the non polarizable models and experimental 
values and in Section 4 the conclusions. \\
\section{The FBA/$\epsilon $ force field  }
\begin{figure}
\centering
\centerline{\psfig{figure=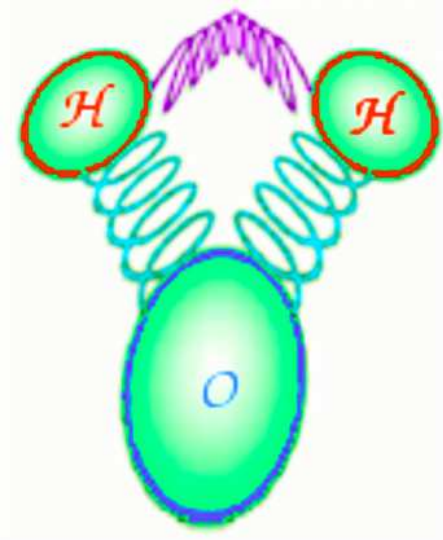,width=6.0cm}}
\caption{model of water including the harmonic potential in bonds and angle.}
\label{model}
\end{figure}

The  Flexible Bond and Angle  (FBA/$\epsilon $) model 
 is illustrated in the Figure~\ref{model}. The 
molecule is represented by
three sites. The oxygen attracts the negative charge while 
the hydrogens have the positive charge. The bonds and angles 
are not fixed but oscillate. The interaction potential exhibits the 
following contributions:
 \begin{equation}
\label{potential}
U(r)=U_{LJ}(r)+U_{e}+U_k(r)+U_{\theta}\; .
\end{equation}

In the Eq.~\ref{potential}, the Lennard-Jones describes
the intermolecular interactions between the 
massive particles, the oxygens, and the 
potential is given by 
\begin{equation}
\label{LJ}
U_{LJ}(r) = 4\epsilon_{\alpha \beta} 
\left[\left(\frac {\sigma_{\alpha \beta}}{r}\right)^{12}-\left (\frac{\sigma_{\alpha \beta}}{r}\right)^6\right]
\end{equation}
where  $r$ is the distance between the
oxygens of two neighbor molecules 
$\alpha$ and $\beta$,  $\epsilon_{\alpha \beta}$ is the LJ energy scale 
and  $\sigma_{\alpha \beta}$ the repulsive 
diameter for an $\alpha \beta$ pair. The cross interactions are obtained 
using the Lorentz-Berthelot mixing rules,
\begin{equation}
\label{lb}
\sigma_{\alpha\beta}= \left(\frac{\sigma_{\alpha\alpha} + \sigma_{\beta\beta} }{2}\right);\hspace{1.0cm} \epsilon_{\alpha\beta}= \left(\epsilon_{\alpha\alpha} \epsilon_{\beta\beta}\right)^{1/2}\;.
\end{equation}
The coulomb forces between oxygen and hydrogen
charges of the same or different
molecules  are represented by
\begin{equation}
\label{e}
U_{e}(r) =  \frac{1}{4\pi\epsilon_0}\frac{q_{\alpha} q_{\beta}}{r}
\end{equation}
where $r$ is the distance between sites $\alpha$ 
and $\beta$, $q_\alpha(\beta)$ 
is the electric charge of site $\alpha(\beta)$
and  $\epsilon_0$ is the permitivity of vacuum.

The difference between the 
rigid SPC/$\epsilon $  and the  FBA/$\epsilon $ model is the 
introduction in the same molecule the intramolecular harmonic potentials in the bonds

 \begin{equation}
\label{k}
U_k(r)=\frac{k_r}{2}(r-r_0 )^2 
\end{equation}
and in the angle.
 \begin{equation}
\label{theta}
U_{\theta}(\theta)=\frac{k_{\theta}}{2}(\theta-\theta_0)^2 ,
\end{equation}
\noindent where $r$ is the bond distance and $\theta$ is the bond 
angle. The subscript $0$ denotes their equilibrium
 values, $k_r$ and $k_{\theta}$ are the corresponding spring constants. 
The model has the following parameters: $\epsilon_{\alpha \beta}$, $\sigma_{\alpha \beta}$, $q_{\alpha}$, $q_{\beta}$, $r_0$, $\theta_0$, $k_r$ and 
$k_{\theta}$. The parameterization procedure is the same
employed in previous publication and goes as follows~\cite{tip4pe}.
 $\epsilon_{\alpha \beta}$, $\sigma_{\alpha \beta}$, $q_{\alpha}$ and $q_{\beta}$ were 
defined by requiring that
the model reproduces: the density
of liquid water, dielectric constant and  dipole moment
at $1\;bar$ and $240 K$ and  the melting temperature at $1\;bar$.
The idea behind is approach is to develop a flexible model
which reproduces the bulk thermodynamic and dynamic
properties of the equivalent rigid model for the pure
system, but due to the flexibility is able to provide
better results in mixtures or in confined geometries.
\begin{table}
\caption{Parameters of the three-site water models considered in this work.
}
\label{table2}
\begin{tabular}{|ccccccccc|}
\hline\hline
model	&	$k_{b}$	&	$r _{OH}$ 	&	$k_{a}$	&	$\Theta$ 	&	$\varepsilon_{OO}$	&	$\sigma_{OO}$	&	$q_{O}$	&	$q_{H}$	\\

	&	kJ/ $mol$ {\AA}$^{2}$ 	&	{\AA}	&	kJ/ $mol$ rad$^{2}$	&	deg	&	kJ mol	&	{\AA}	&	e	&	e	\\
SPC/E \cite{spce}	&	-	&	1.000	&	-	&	109.45	&		&	3.1660	&	-0.8476	&	0.4238	\\
SPC/$\epsilon$	\cite{spce}&	-	&	1.000	&	-	&	109.45	&		&	3.1785	&	-0.8900	&	0.4450	\\
FBA/$\epsilon $	&	3000	&	1.027	&	383	&	114.70	&	0.792324	&	3.1776	&	-0.8450	&	0.4225	\\
SPC/Fw ~\cite{Wu}	&	4236.648	&	1.012	&	317.56	&	113.24	&	0.650299	&	3.165492	&	-0.8200	&	0.4100	\\
\hline
\end{tabular}
\end{table}
\section{Simulation Details}
We performed molecular dynamics simulations in the isothermal-isobaric
ensemble, NPT, with isotropic fluctuations of volume, to compute 
the liquid properties at different temperatures and
 standard pressure, $1\;bar$. These simulations involved
typically 500 molecules.

In order to compute the surface tension we used the
constant volume and temperature ensemble,
NVT, and   5832 molecules. We obtained
the  liquid-vapor interface
 by setting up a liquid slab surrounded by vacuum
in a simulation box with periodic boundary conditions in the
three spatial directions. The dimensions of the simulation cell
were $Lx = Ly = 54$ {\AA} with $Lz = 3Lx$, with $z$ being the normal
direction to the liquid-vapor interface.
The GROMACS 4.5.4 package~\cite{Hess, Spoel} was employed in all 
simulations presented in this work. The equations of motion
were solved using the leapfrog algorithm with a time step of $1\;fs$. The
 temperature was coupled to the Nos\'e-Hoover
thermostat with a parameter ${\tau_T}= 0.2\;ps$ while the pressure
was coupled to the Parrinello-Rahman barostat~\cite{Parrinello}with a
coupling parameter ${\tau_P}= 0.5\;ps$. 

We computed the  electrostatic interactions
 with the particle mesh Ewald approach~\cite{Essmann} with a
tolerance of $10^{6}$ for the real space contribution, with a grid
spacing of $1.2$ {\AA} and spline interpolation of order 4. In the
isotropic NPT simulations the real part of the Ewald
summation and the LJ interactions were truncated at $9$ {\AA}.
Long range corrections for the LJ energy and pressure were
included. The dielectric constant is obtained from the analysis of 
the dipole moment fluctuations of the simulation 
system~\cite{Neumann, Hansen}. The density and the dielectric constant
 were calculated from the
same simulation for at least $200\;ns$ after an equilibration period
of $10\;ns$. For the surface tension computations in the NVT
ensemble the cutoff was set to $26$ {\AA}, since the surface tension
depends on the truncation of the interactions \cite{Truckymchuk} and the
interface cross-sectional area.\cite{Orea, minerva} The equilibration 
period for the interfacial simulations was $2\; ns$, and the results for the
average properties were obtained over an additional $10\; ns$
trajectory. 

The Berensend barostat was employed
For the calculation of the melting temperature and of the  density of 
the ice. The use of this
 barostat allows the simulation box to expand or contract, 
and then  to form ice or liquid phases. For 
studying the ice phase and the 
melting properties,  the temperature was fixed with a Berendsen thermostat 
 with a relaxation time of $0.2\;ps$~\cite{Garcia}.  For the description of 
the coexistence between liquid and solid water, we employed an 
orthogonal cell. This approach is consistent with the 
crystallographic data of the solid phase Ih~\cite{Petrenko}. The cell size 
is $Lx =21.6${\AA}, $Ly=23.3${\AA} and $Lz=53.8${\AA} .Which gives us a 
contact area between the $Lx*Ly=503.28${\AA}${^2}$ phases. 
\section{Results}
\begin{figure}
\centerline{\psfig{figure=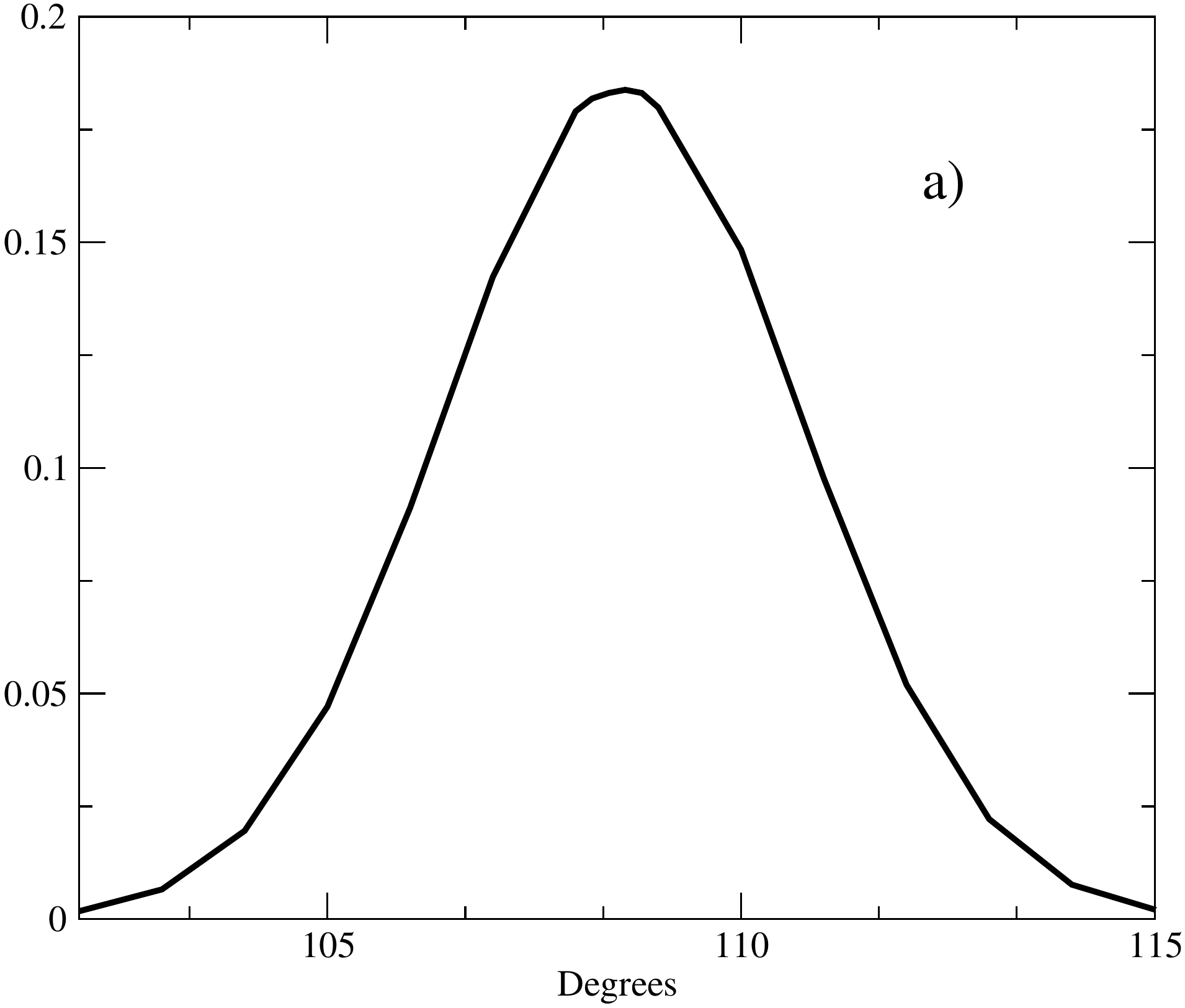,width=8.0cm}
\psfig{figure=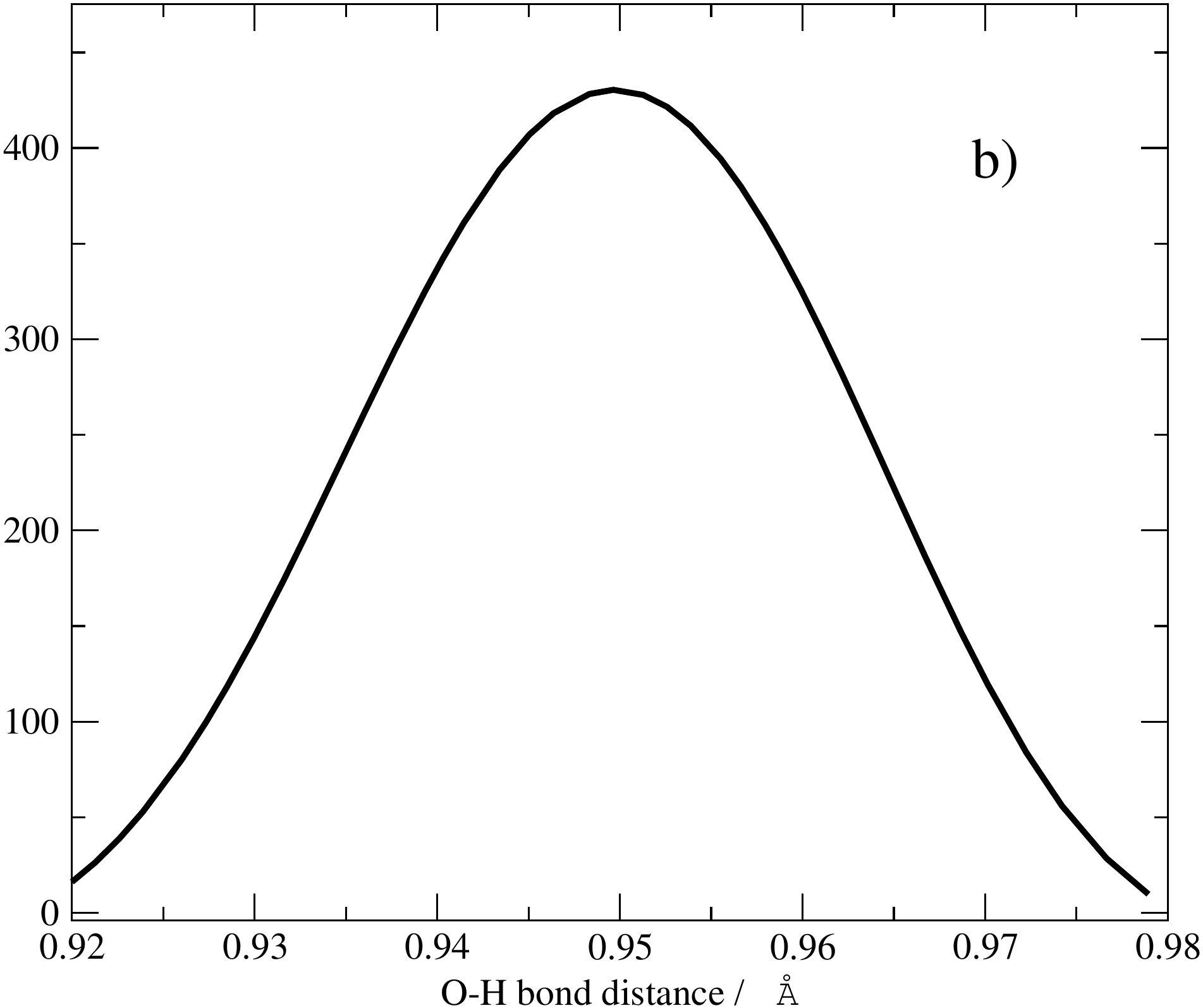,width=8.0cm}}
\centerline{
\psfig{figure=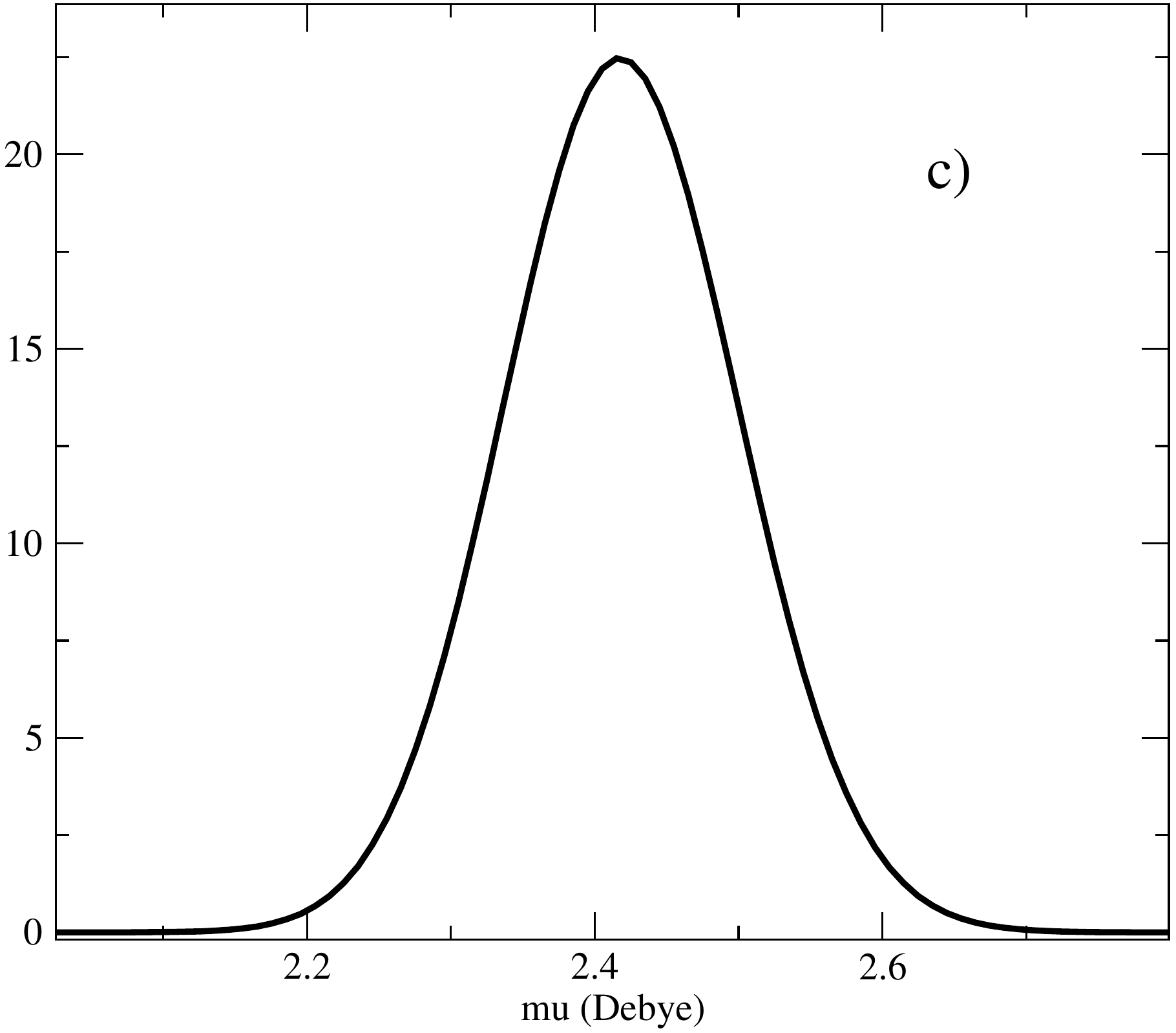,width=8.0cm}}
\caption{(a)  HOH angle (b)O-H bond distance (c) dipole moment distributions at 298K and 1bar for the  FBA/$\epsilon $ model. }
\label{angD}
\end{figure}
First, we analyzed the  water  structure obtained using our model..
Differently from the rigid models, the FBA/$\epsilon $ exhibits
a distribution of  HOH angles illustrated for $1\; bar$ and $298\;K$ in 
the figure~\ref{angD}(a). The average angle, $108.56^o$ is
close to the average experimental value~\cite{Ichikawa} which
is $106^o$ and to the value employed for the 
 rigid  SPC model which is  $109.47^o$. 
The distribution of  O-H bond distances for 
the FBA/$\epsilon $ model at $298\; K$ and $1\;bar$ is illustrated  in the 
figure~\ref{angD}(b). This result shows the average bond distance at 
of $0.09495\;nm$ what
is $4\%$ lower than the  neutron diffraction 
value, $0.099\;nm$~\cite{Zeidler}, and
only  $2.4\%$ lower than the X-ray diffraction 
value, $ 0.09724\;nm$~\cite{Na71,To00}. In principle
rigid models can be constructed to give this bond distance, however 
they can not accommodate the change with the temperature
of the O-H bond distance observed both in the experiments and 
in our model. Figure~\ref{angD}(c) shows the  distribution of dipole moments
of  the water molecules at $298\;K$ and $1$ bar.
The  mean dipole moment of the distribution
is $2.42\;D$. 
\begin{figure}
\centering {
\psfig{figure=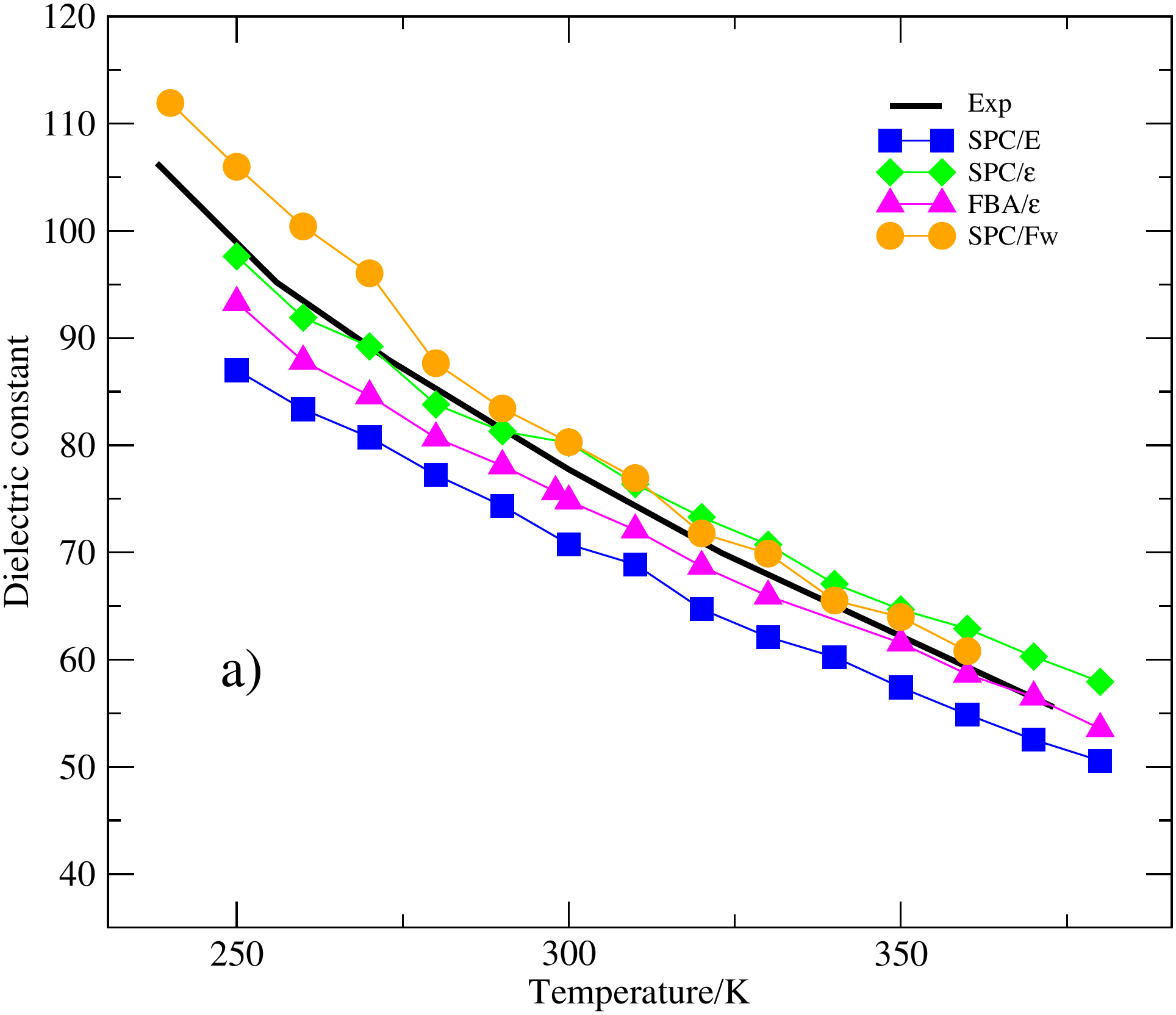,width=8.0cm}
\psfig{figure=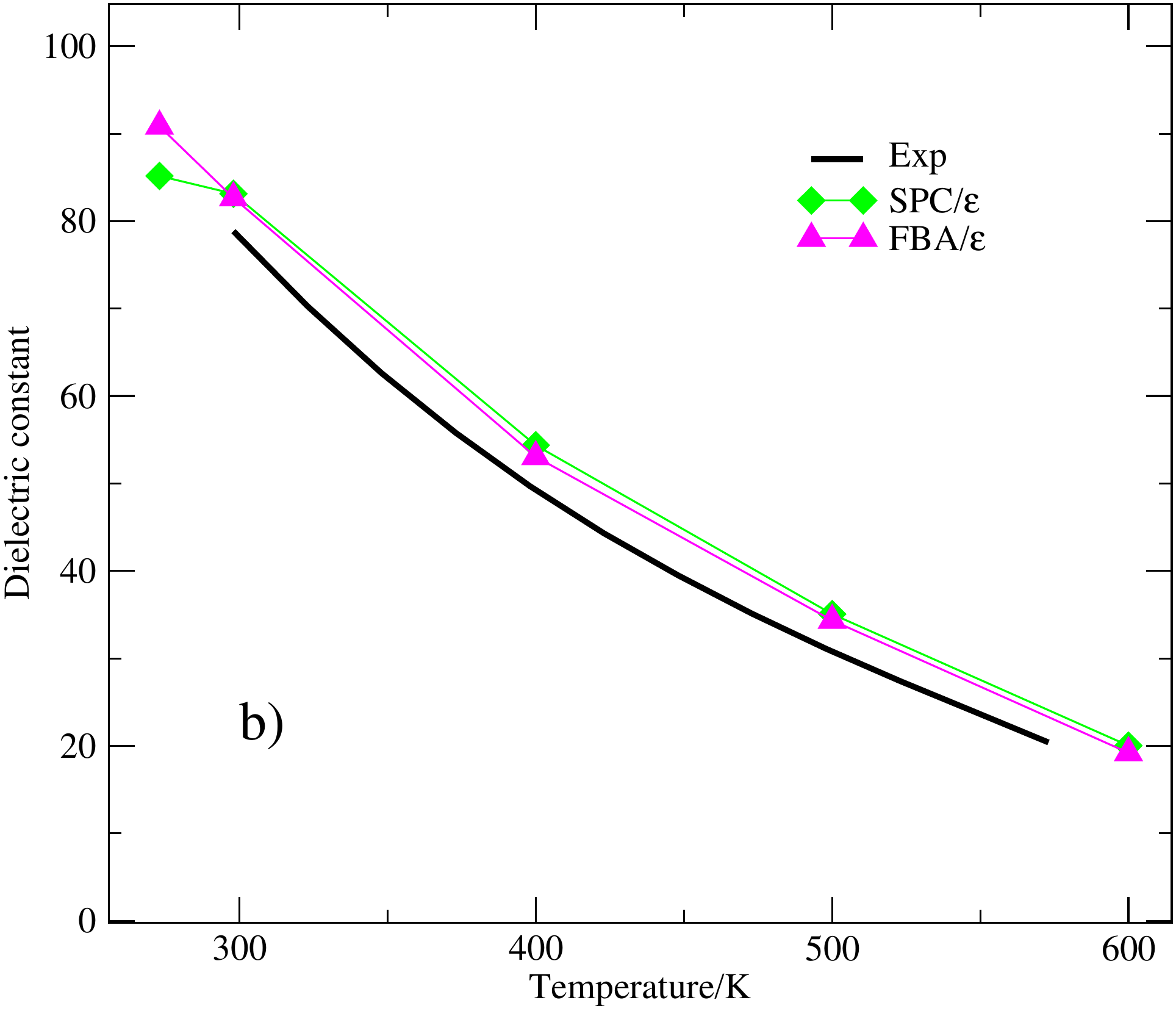,width=8.0cm}}
\centering {
\psfig{figure=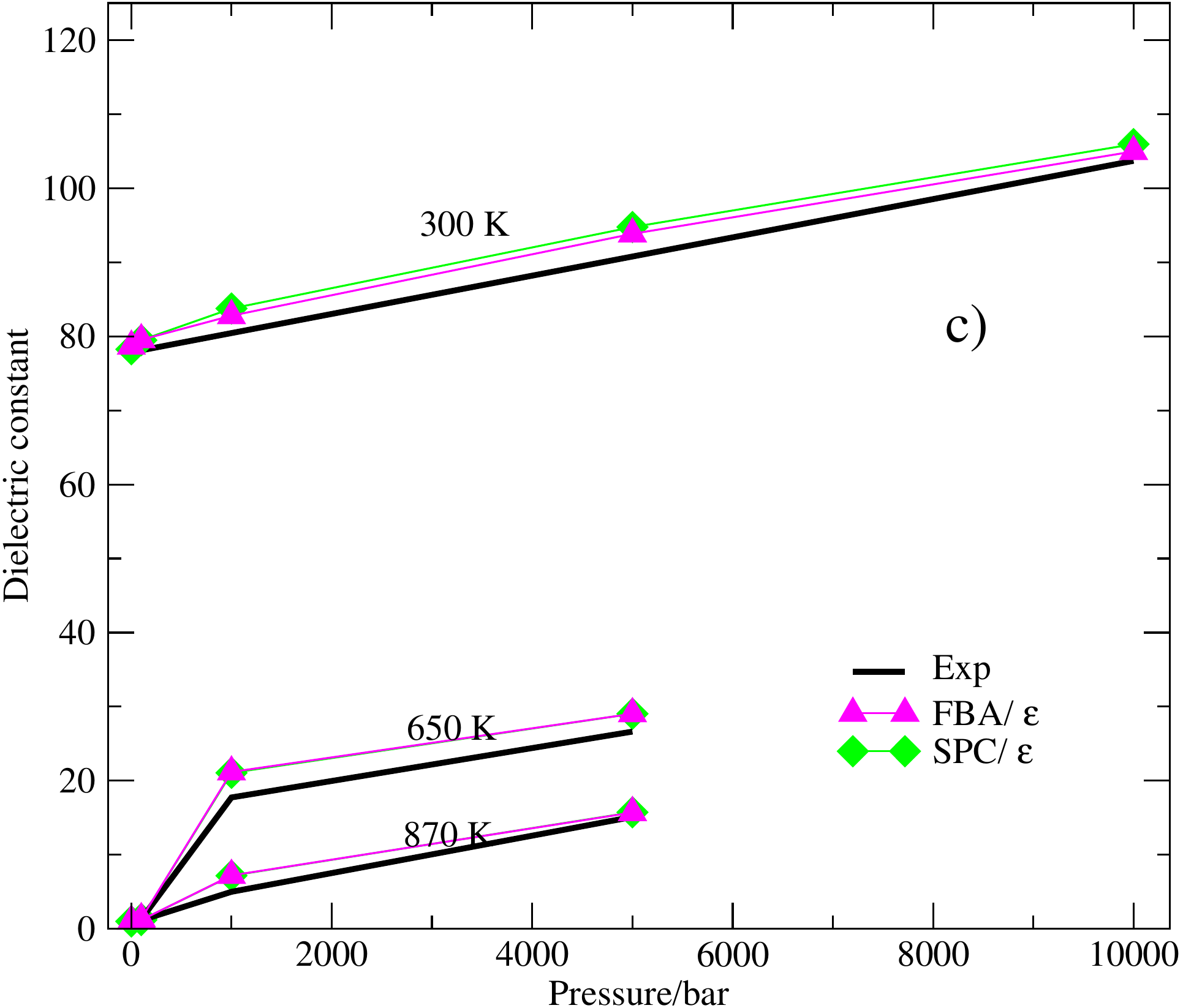, width=8.0cm}}
\caption{Dielectric constant  (a) versus temperature  at $1$ bar,  (b) at the liquid-vapor coexistence line and (c) versus pressure for three different temperatures
SPC/E~\cite{spce}(squares), SPC/$\epsilon$~\cite{spce}(diamonds), FBA/$\epsilon$ (triangles), SCP/Fw calculated here (circles) models and experimental data~\cite{NIST} (solid line).}
\label{Eps-T}	
\end{figure}
Then, we test  how robust is the parametrization
regarding variations of temperature.
The flexible  FBA/$\epsilon $ and
the SPC/$\epsilon $  models are  parametrized to 
reproduce  the experimental value of the 
dielectric constant, $\epsilon$, at $298\;K$ and $1\;bar$ with a $3.6\;\%$
of tolerance~\cite{fernandez}.
 Figure~\ref{Eps-T} illustrates the dielectric constant,
$\epsilon$, versus pressure
for different temperatures at $1$ bar and at the 
liquid-vapor coexistence for both models. The comparison
between the  FBA/$\epsilon $, the rigid model and 
the experiments indicates that flexibility does
not adds better concordance with the data even if 
large changes is pressure and in temperature are 
implemented as shown in the figure~\ref{Eps-T}(c).

\begin{figure}
\centering
\centerline{\psfig{figure=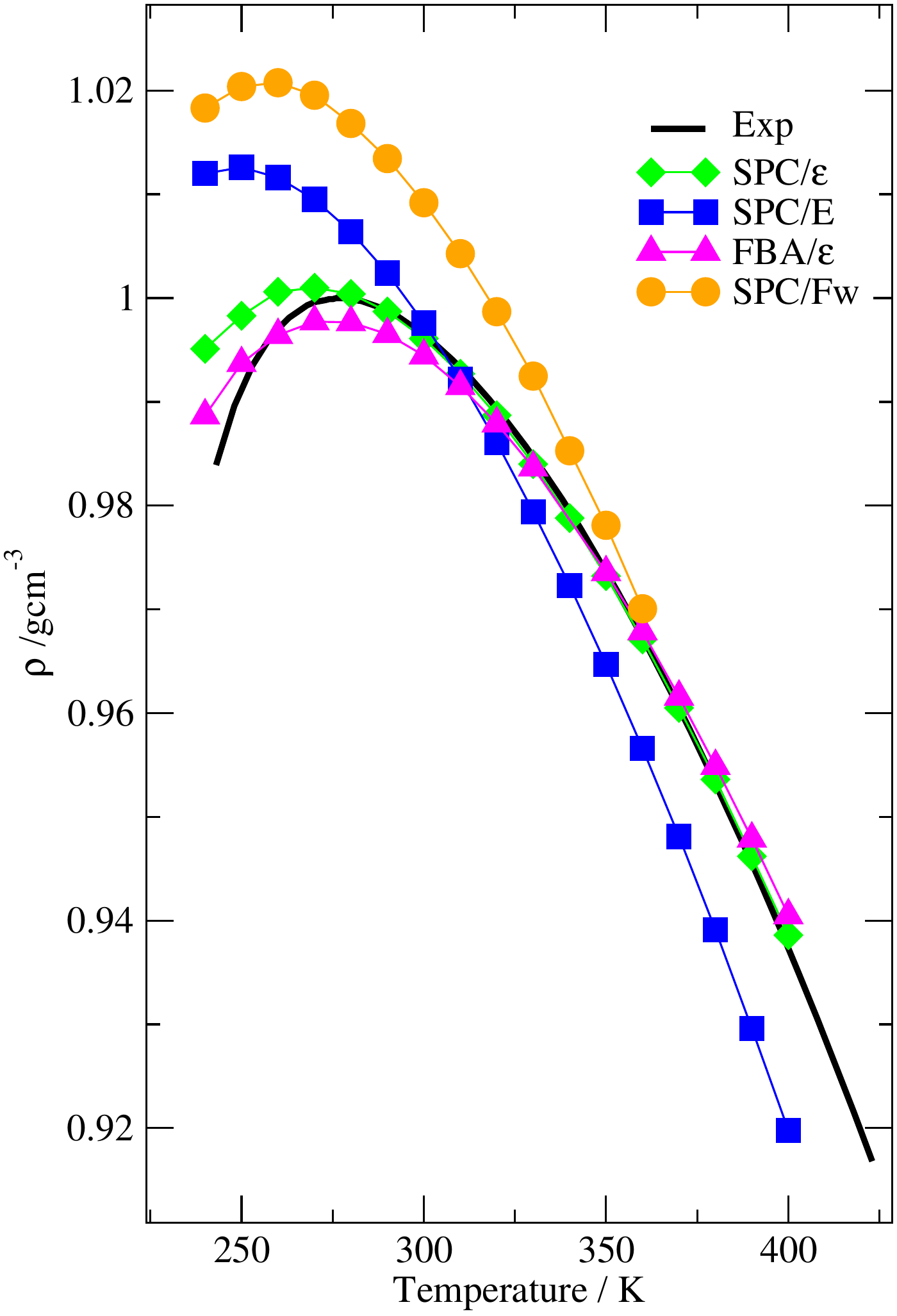,width=8.0cm}}
\caption{Density as a function of temperature at 1 bar for the
SPC/E~\cite{spce}(squares), SPC/$\epsilon$~\cite{spce}(diamonds), FBA/$\epsilon$ (triangles), SCP/Fw calculated here (circles) models and experimental data~\cite{NIST} (solid line).}
\label{dens}	
\end{figure}

The FBA/$\epsilon $ model was
also parametrized to reproduce the experimental density 
at  $298\;K$ and $1\;bar$~\cite{NIST}.
The  figure~\ref{dens} 
shows the density as a function of the temperature for 
the FBA/$\epsilon $,   SPC/E~\cite{spce}, SPC/$\epsilon $~\cite{spce}, SPC/$Fw$~\cite{Wu}
 models and 
the experiments~\cite{NIST}. The non-polarizable models and the FBA/$\epsilon $ agree with
the experiments at $300K$ since they were parametrized to
give the correct density at this temperature. The  SPC/E~\cite{spce},
however, at low temperatures overestimates the density, while both
the  SPC/$\epsilon $~\cite{spce} and FBA/$\epsilon $ agree with
the experiments for a wide range of temperatures. The FBA/$\epsilon $
at very low temperatures shows a small improvement over the  
SPC/$\epsilon $~\cite{spce} model.
\begin{figure}
\centerline{
\psfig{figure=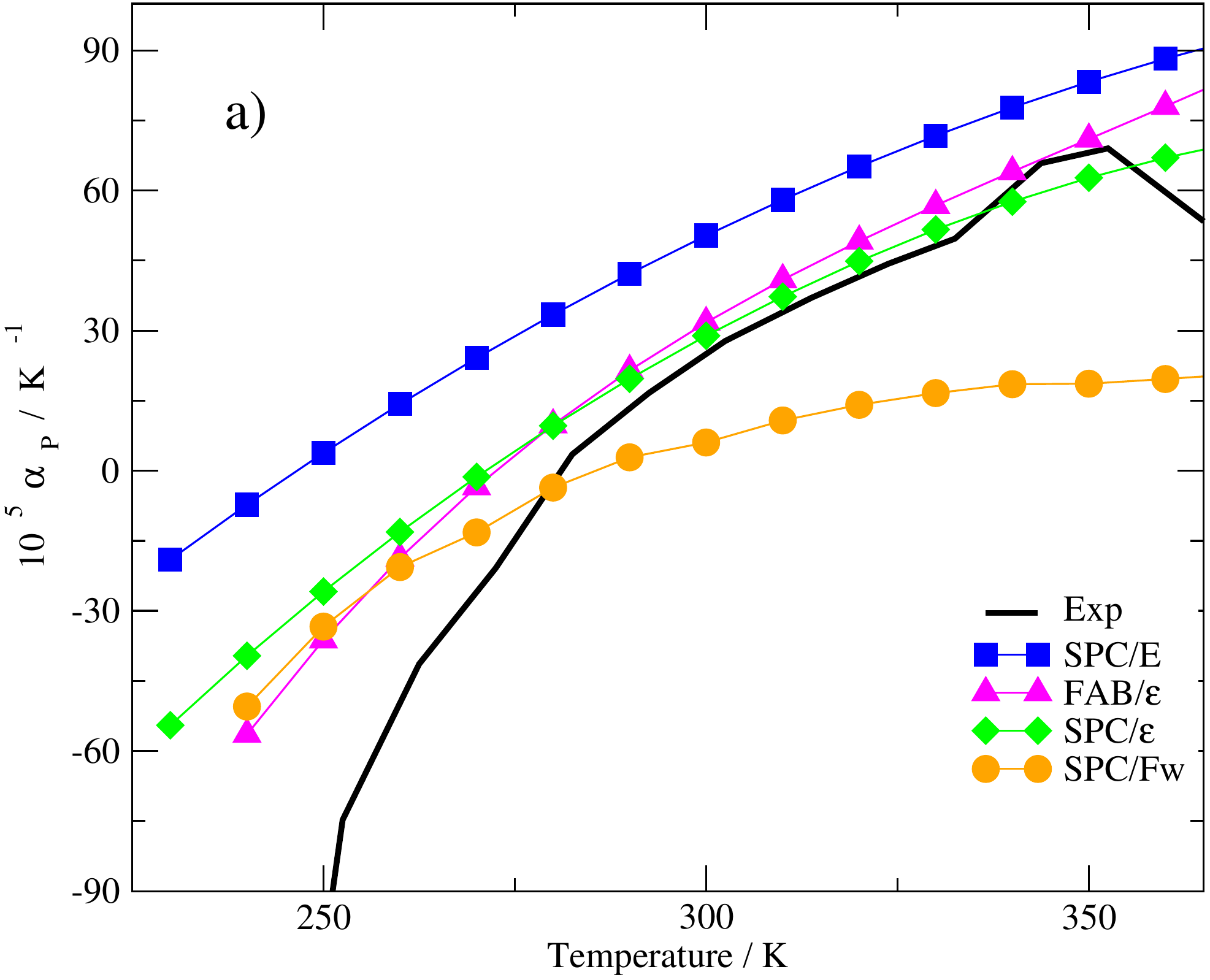,width=8.0cm}
\psfig{figure=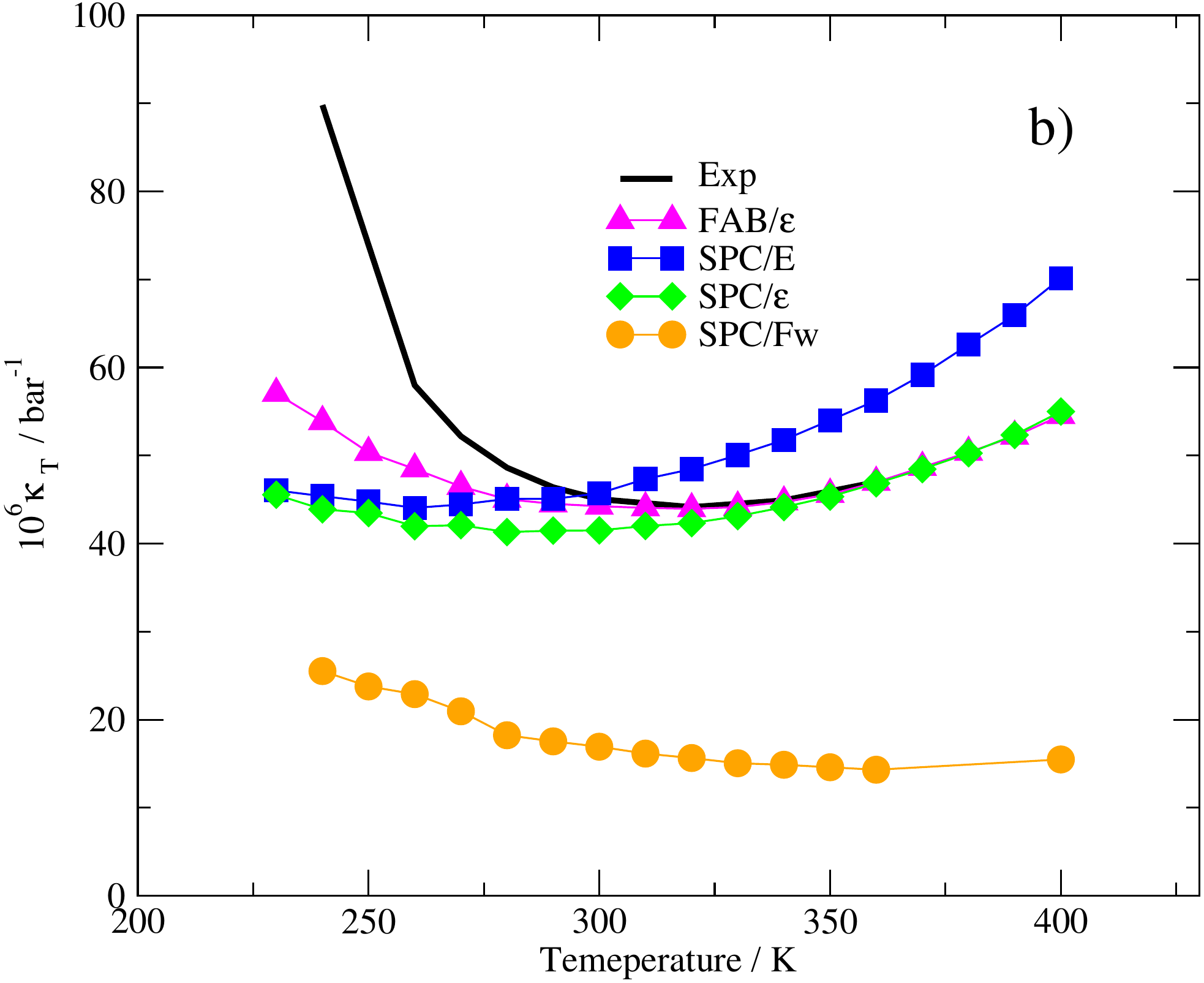,width=8.0cm}}
\caption{(a) Thermal expansion coefficient and (b) isothermal compressibility  as a function of temperature at pressure constant of 1bar for the
SPC/E~\cite{spce}(squares), SPC/$\epsilon$~\cite{spce}(diamonds), FBA/$\epsilon$ (triangles), SCP/Fw calculated here (circles) models and experimental data~\cite{NIST} (solid line).}
\label{alfa}
\end{figure}
A consequence of a good
parametrization of the density for a wide range of temperatures can be
observed in the behavior of the 
 response functions.
Response functions exhibit a very  peculiar behavior 
in water. The thermal expansion coefficient, $\alpha$, which for most
materials is almost
constant or slightly increases  with the temperature, for water it decreases 
abruptly with the decrease of the temperature
and it becomes negative. The compressibility for a number of materials
increases monotonically  with the temperature but in the  case of 
water it has a minimum. Figure~\ref{alfa} illustrates
both $\alpha$ and $\kappa_T$ as 
a function of temperature  for  SPC/E, SPC/$\epsilon$,
  FBA/$\epsilon$ models 
and the experimental results~\cite{NIST}.  At low temperatures the FBA/$\epsilon$ model
gives the better approximation to the experimental data~\cite{NIST}. In the particular 
case of the compressibility, the flexible model
shows the minimum approximately in the same temperature as the experiments 
while the non-polarizable force fields present a shift.
\begin{figure}
\centerline{\psfig{figure=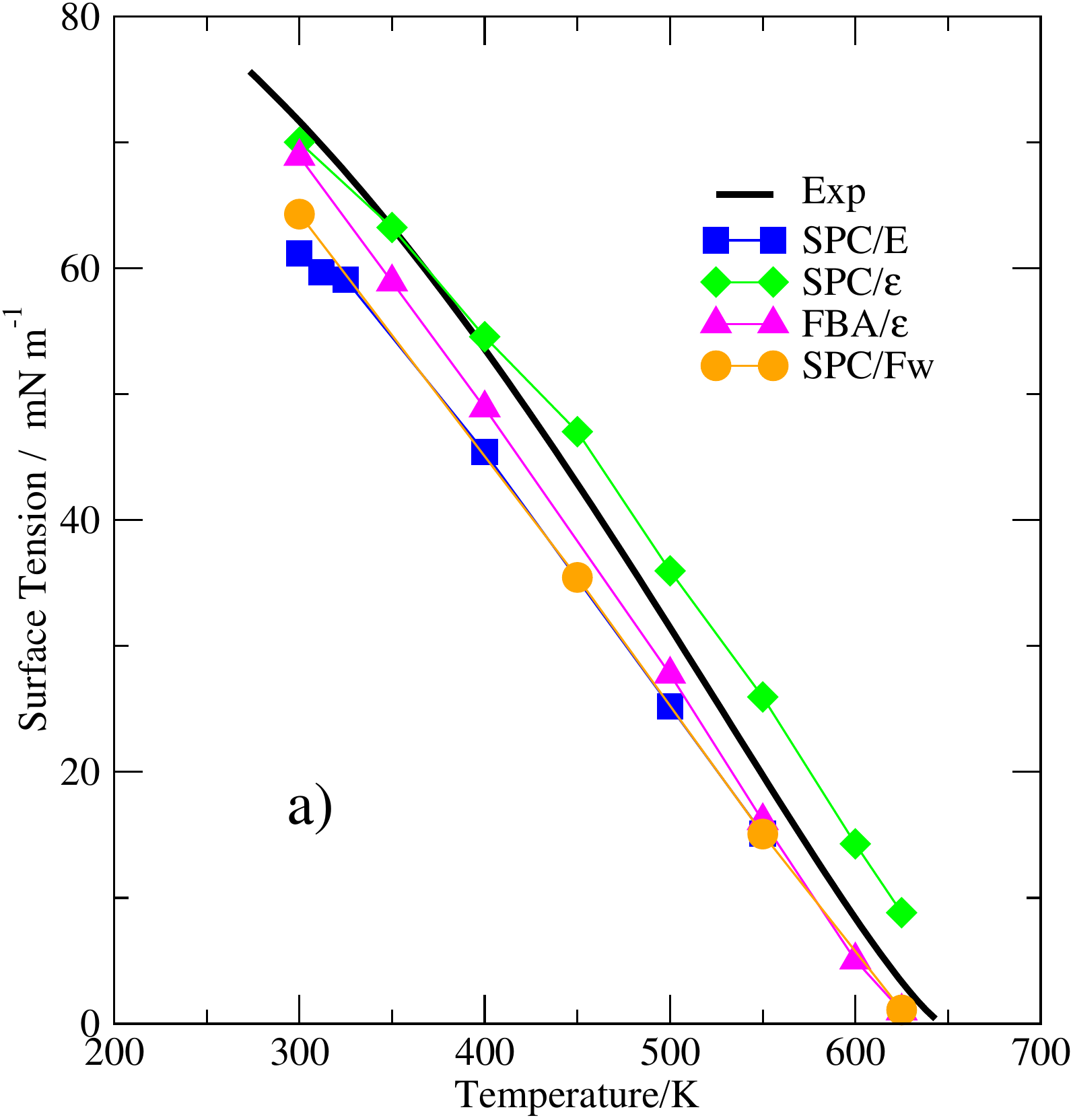,width=7.0cm}
\psfig{figure=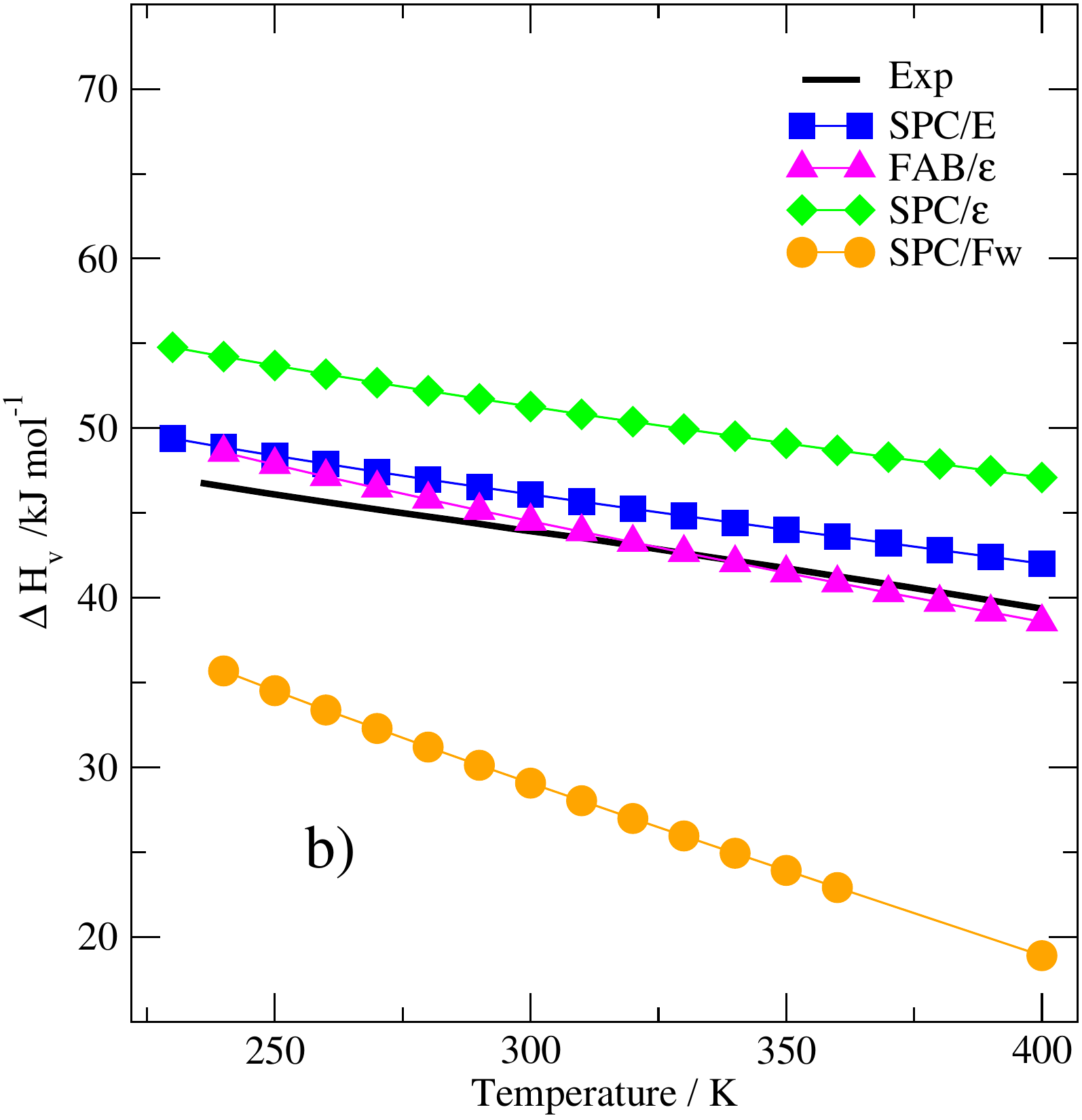,width=7.0cm}
}
\caption{(a)Surface tension and (b) heat of vaporization 
as a function of temperature at pressure constant of 1bar for the
SPC/E~\cite{spce}(squares), SPC/$\epsilon$~\cite{spce}(diamonds), FBA/$\epsilon$ (triangles), SCP/Fw calculated here (circles) models and experimental data~\cite{NIST} (solid line).}
\label{stens}
\end{figure}
\begin{figure}
\centerline{\psfig{figure=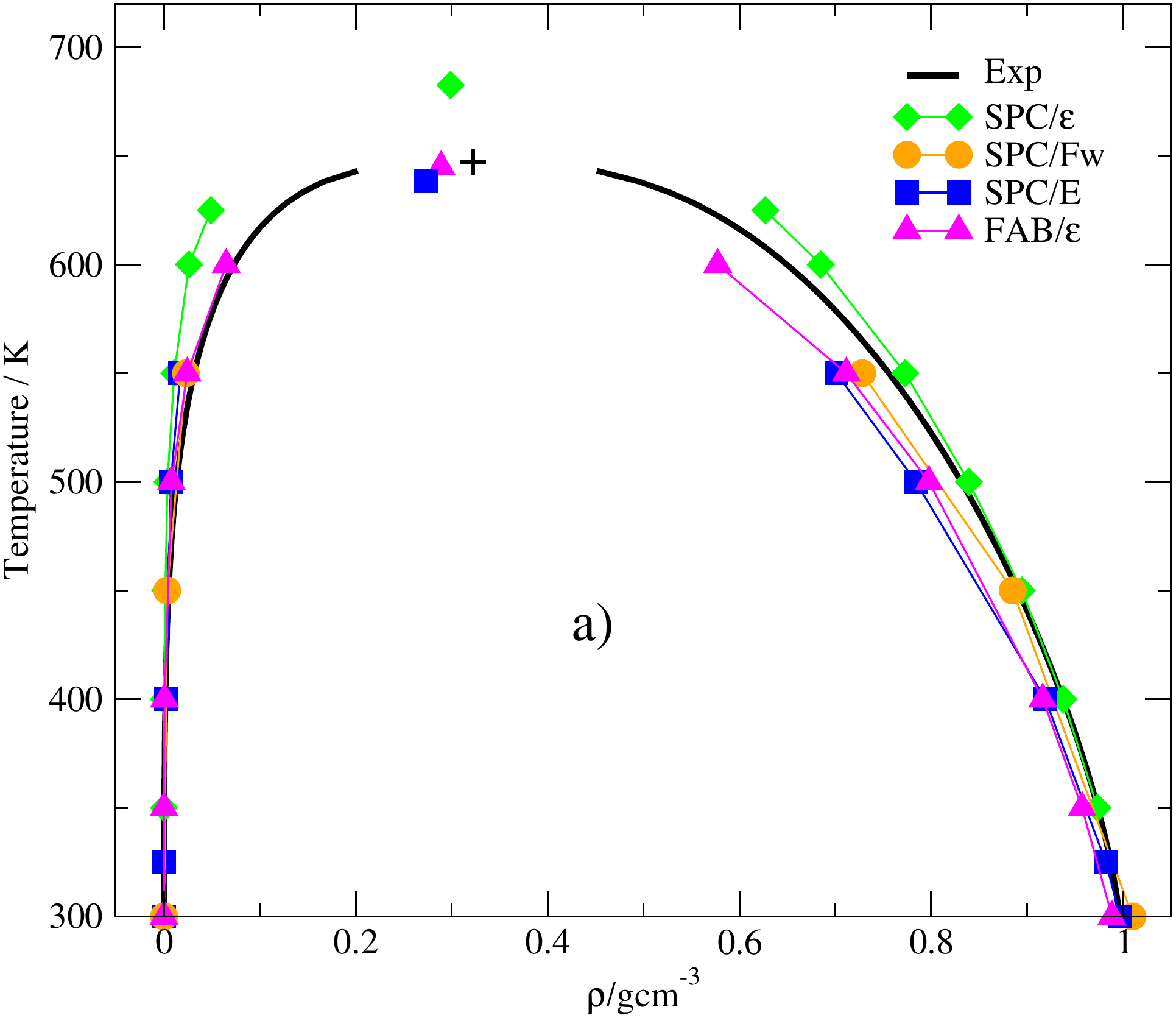,width=7.0cm}
\psfig{figure=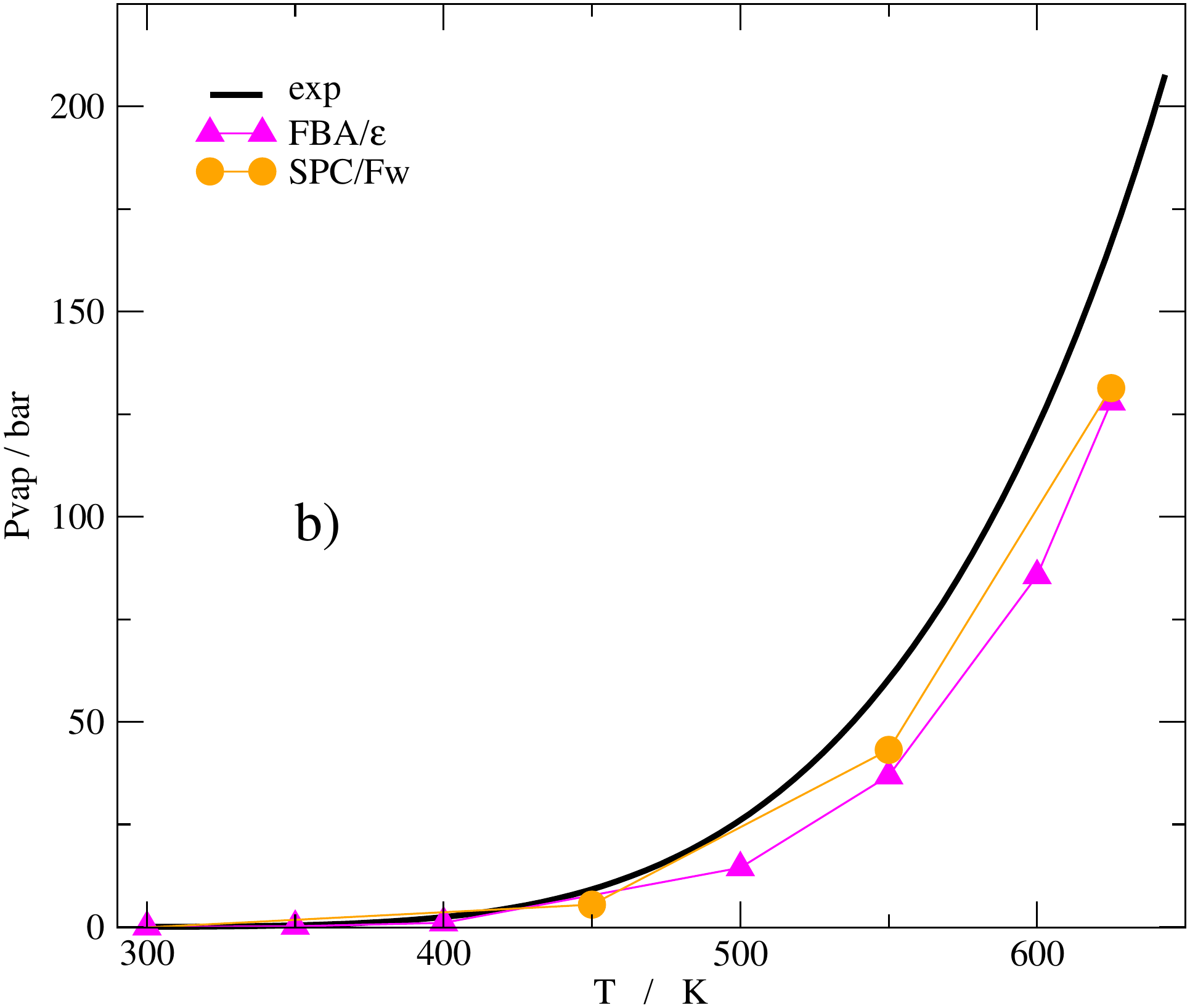,width=7.0cm}}
\caption{ Temperature versus density phase diagram for the SPC/E~\cite{spce}(squares), SPC/$\epsilon$~\cite{spce}(diamonds), FBA/$\epsilon$ (triangles), SCP/Fw calculated here (circles) models and experimental data~\cite{NIST} (solid line).}
\label{LV}
\end{figure}
 Next we check if the flexible model also presents a
good agreement with the experiments for high temperatures
at the vapor phase given the parametrization was performed
at the low temperatures and densities. 
One important property in which the flexibility might matter
is the surface tension. For testing the behavior at high 
temperatures we selected the surface tension.
The surface tension versus temperature is shown
in the figure~\ref{stens}(a). The flexible model shows a consistent
agreement with the experiment while the rigid models 
fit the data either at low or high temperatures but not for both ranges.
As the temperature is 
increased, it is not clear how the   HOH angle and the  OH  bond might change.
 In order to check this we look at the liquid-vapor transition. Figure~\ref{stens}(b)
 compares  the heat
of vaporization as a function of the temperature at $1\;bar$ 
for the 
SPC/E~\cite{spce}(squares), SPC/$\epsilon$~\cite{spce}(solid
diamonds), FBA/$\epsilon$ (triangles) models with the 
experimental data~\cite{NIST} (solid line). It shows that
 both rigid SPC/$\epsilon$ and 
flexible FBA/$\epsilon$ agree with the data what suggests that
the flexibility is not so relevant for the vapor heat specifically
but might affect coexistence properties. In order to check
this point,  we also obtain the gas-liquid phase diagram. 
The coexisting densities were estimated from the average 
density profile in the liquid and vapor regions of the 
slab.  Figure~\ref{LV}(a)
 shows the temperature versus density 
phase diagram for the
SPC/E~\cite{spce}(squares), SPC/$\epsilon$~\cite{spce}(diamonds), FBA/$\epsilon$ (triangles)
 models and experimental data~\cite{NIST} (solid line). The flexible model 
improves the agreement with 
experiment significantly compared to the
rigid  models for the gas phase and for
the critical temperature and density.
The vapor pressure  is calculated as the normal
component of the pressure tensor in the interface simulations.
 Consistent with the improved results
for the gas phase, figure~\ref{LV}(b) shows
that the new force field reproduces the curvature of 
vapor pressure when compared with the experiments~\cite{NIST}.

Finally, a comprehensive comparison between the results
obtained by our flexible model compared with
its non-polarizable  approaches is presented in the Table~\ref{table-calif3}.
Following the notation introduced by Vega et al.~\cite{vega11}, the
models are evaluate by a score. The final score of the
FBA/$\epsilon$ is higher than the non-polarizable  models.
\begin{table}
\caption{Experimental and simulation data results of 3 sites water models. The thermodynamic conditions are reported according to the calculated property.}
\scalebox{0.7}[0.65]{
\begin{tabular}{|l|c|c|c|c|c|c|c|c|}
\hline
	&		&		&		&		&		&		&		&		\\
Property	&	Experimental	&	SPC/E	&	SPC/$\varepsilon$	&	FBA/$\varepsilon$	&	Tolerance ($\%$)	&	SPC/E	&	SPC/$\varepsilon$	&	FBA/$\varepsilon$	\\
	&	data	&		&		&		&		&	score	&	score	&	score	\\
\hline																	
\multicolumn{9}{|l|}{Enthalpy of phase change / kcal mol$^{-1}$}\\																	
\hline																	
$\Delta$H$_{melt}$	&	1.44	&	0.74	&	5.38	&	2.14	&	5	&	0.3	&	0.0	&	0.3	\\
$\Delta$H$_{vap}$	&	10.52	&	11.79	&	12.25	&	10.69	&	2.5	&	5.2	&	3.4	&	9.4	\\
\hline																	
\multicolumn{9}{|l|}{Critical point properties}\\																	
\hline																	
T$_C$/K	&	647.1	&	638.6	&	682.6	&	642	&	2.5	&	9.5	&	7.8	&	9.7	\\
$\rho_C$/g cm${^{-3}}$	&	0.322	&	0.273	&	0.299	&	0.299	&	2.5	&	3.9	&	7.1	&	7.1	\\
p$_C$/bar	&	220.64	&	139	&	167	&	167	&	5	&	2.6	&	5.1	&	5.1	\\
\hline																	
\multicolumn{9}{|l|}{Surface tension/mN m$^{-1}$}\\																	
\hline																	
$\gamma_{300K}$	&	71.73	&	63.6	&	70.02	&	73.98	&	2.5	&	5.5	&	9.0	&	8.7	\\
$\gamma_{450K}$	&	42.88	&	36.7	&	47	&	46.2	&	2.5	&	4.2	&	6.2	&	6.9	\\
\hline																	
	\multicolumn{9}{|l|}{Melting properties}\\																
\hline																	
T$_m$/K	&	273.15	&	215	&	200	&	243	&	2.5	&	1.5	&	0.0	&	5.6	\\
$\rho_{liq}$/g cm$^{-3}$	&	0.999	&	1.011	&	0.9864	&	0.9901	&	0.5	&	7.6	&	7.5	&	8.2	\\
$\rho_{sol}$/g cm$^{-3}$	&	0.917	&	0.95	&	0.8932	&	0.945	&	0.5	&	2.8	&	4.8	&	3.9	\\
dp/dT (bar K$^{-1}$)	&	-137	&	-126.05	&	-591.52	&	-424.982	&	5	&	8.4	&	0.0	&	0.0	\\
\hline																	
\multicolumn{9}{|l|}{Orthobaric densities and temperature of maximun density \textbf{TMD}}\\																	
\hline																	
\textbf{TMD}/K	&	277	&	241	&	266	&	275.4	&	2.5	&	4.8	&	8.4	&	9.8	\\
$\rho_{298K}$/g cm$^{-3}$	&	0.997	&	0.994	&	0.9964	&	0.9948	&	0.5	&	9.4	&	9.9	&	9.6	\\
$\rho_{400K}$/g cm$^{-3}$	&	0.9375	&	0.916	&	0.9385	&	0.9406	&	0.5	&	5.4	&	9.8	&	9.3	\\
$\rho_{450K}$/g cm$^{-3}$	&	0.8903	&	0.86	&	0.8893	&	0.8982	&	0.5	&	3.2	&	9.8	&	8.2	\\
\hline																	
	\multicolumn{9}{|l|}{Isothermal compressibility / 10$^-6$ bar$^{-1}$)}\\																
\hline																	
$\kappa_T$ [1 bar; 298 K]	&	45.3	&	46.1	&	41.4	&	45.6	&	5	&	9.6	&	8.3	&	9.9	\\
$\kappa_T$ [1 bar;360 K]	&	47	&	57.7	&	46.86	&	47.2	&	5	&	5.4	&	9.9	&	9.9	\\
\hline																	
\multicolumn{9}{|l|}{Thermal expansion coefficient   / 10$^5$ K$^{-1}$)}\\																	
\hline																	
$\alpha_P$ [1 bar; 298 K]	&	22.66	&	48.6	&	26.9	&	29.5	&	5	&	0.0	&	6.3	&	4.0	\\
$\alpha_P$ [1 bar;350 K]	&	68.2	&	83.23	&	62.56	&	71	&	5	&	5.6	&	8.3	&	9.2	\\
\hline																	
\multicolumn{9}{|l|}{Gas properties}\\																	
\hline																	
$\rho_{v}$[350 K] (bar)	&	0.417	&	0.14	&	0.042	&	0.224	&	5	&	0.0	&	0.0	&	0.7	\\
$\rho_{v}$[450 K] (bar)	&	9.32	&	5.8	&	1.88	&	5	&	5	&	2.4	&	0.0	&	0.7	\\
\hline																	
\multicolumn{9}{|l|}{Heat capacity at constant pressure/cal mol$^{-1}$K$^{-1}$}\\																	
\hline																	
C$_p$[liq 298 K; 1 bar]	&	18	&	20.7	&	20.6	&	20.7	&	5	&	7.0	&	7.1	&	7.0	\\
C$_p$[ice 250 K; 1 bar]	&	8.3	&	14.9	&	14.8	&	14.9	&	5	&	0.0	&	0.0	&	0.0	\\
\hline																	
\multicolumn{9}{|l|}{Static dielectric constant}\\																	
\hline																	
$\varepsilon$[liq; 298 K]	&	78.5	&	68	&	78.3	&	75.5	&	2.5	&	4.6	&	9.9	&	8.5	\\
$\varepsilon$[liq; 350 K]	&	62.12	&	57.45	&	64.65	&	61.49	&	2.5	&	7.0	&	8.4	&	9.6	\\
$\varepsilon$[10kbar,300K]	&	103.63	&	91.3	&	106.65	&	104.9	&	2.5	&	5.2	&	8.8	&	9.5	\\
$\varepsilon$[I$_h$; 240 K]	&	107	&	39	&	23	&	39.5279	&	2.5	&	0.0	&	0.0	&	0.0	\\

\hline																	
\multicolumn{9}{|l|}{T$_m$-TMD-T$_c$. ratios}\\																	
\hline																	
T$_m$[I$_h$]/T$_c$	&	0.422	&	0.337	&	0.286	&	0.378	&	5	&	6.0	&	3.6	&	7.9	\\
TMD/T$_c$	&	0.428	&	0.378	&	0.381	&	0.428	&	5	&	7.7	&	7.8	&	10.0	\\
TMD-T$_m$(K)	&	4	&	26	&	66	&	32.4	&	5	&	0.0	&	0.0	&	0.0	\\

\hline																	
\multicolumn{9}{|l|}{Densities of ice polymorphs/g cm$^{-3}$}\\																	
\hline																	
$\rho$[I$_h$ 250 K; 1 bar]	&	0.92	&	0.944	&	0.907	&	0.94	&	0.5	&	4.8	&	7.2	&	5.7	\\
$\rho$[II 123 K; 1 bar]	&	1.19	&	1.245	&	1.18	&	1.245	&	0.5	&	0.8	&	8.3	&	0.8	\\
$\rho$[V 223 K; 5.3 kbar]	&	1.283	&	1.294	&	1.273	&	1.294	&	0.5	&	8.3	&	8.4	&	8.3	\\
$\rho$[VI 225 K; 11 kbar]	&	1.373	&	1.403	&	1.33	&	1.403	&	0.5	&	5.6	&	3.7	&	5.6	\\
\hline																	
\multicolumn{9}{|l|}{EOS high pressure}\\																	
\hline																	
$\rho$[373 K; 10 kbar]	&	1.201	&	1.213	&	1.2034	&	1.215	&	0.5	&	8.0	&	9.6	&	7.7	\\
$\rho$[373 K; 20 kbar]	&	1.322	&	1.338	&	1.3219	&	1.339	&	0.5	&	7.6	&	10.0	&	7.4	\\
\hline																	
\multicolumn{9}{|l|}{Self-diffusion coefficient/cm$^2$s$^{-1}$ }\\																	
\hline																	
ln D$_{278K}$	&	-11.24	&	-11.08	&	-11.69	&	-11.58	&	0.5	&	7.2	&	2.0	&	4.0	\\
ln D$_{298K}$	&	-10.68	&	-10.58	&	-11.08	&	-11.01	&	0.5	&	8.1	&	2.5	&	3.8	\\
ln D$_{318K}$	&	-10.24	&	-10.24	&	-10.72	&	-10.71	&	0.5	&	10.0	&	0.6	&	0.8	\\
E$_a$ k$_J$ mol$^{ -1}$	&	18.4	&	15.4	&	17.82	&	15.98	&	5	&	6.7	&	9.4	&	7.4	\\
\hline																	
\multicolumn{9}{|l|}{Shear viscosity / mPa s}\\																	
\hline																	
$\eta$[1 bar; 298 K]	&	0.896	&	0.729	&	1.259	&	0.9443	&	5	&	6.3	&	1.9	&	8.9	\\
$\eta$[1 bar; 373 K]	&	0.284	&	0.269	&	0.378	&	0.3259	&	5	&	8.9	&	3.4	&	7.0	\\
\hline																	
\multicolumn{9}{|l|}{Orientational relaxation time / ps}\\													
\hline																	
$\tau_2^{HH}$ [1 bar; 298 K]	&	2.36	&	1.9	&	1.97	&	2.1	&	5	&	6.1	&	6.7	&	7.8	\\
\hline	
\multicolumn{9}{|l|}{Structure}\\	
\hline																
$\chi^2$(F(Q))	&	0	&	17.7	&	17.2	&	18.1	&	5	&	8.0	&	8.0	&	8.0	\\
$\chi^2$(overall)	&	0	&	22.2	&	21.9	&	22.6	&	5	&	7.0	&	7.0	&	7.0	\\
\hline																	
\multicolumn{6}{|l|}{Phase diagram}			&	2.0	&	2.0	&	7.0	\\
\hline																	
																	
\hline																	
\multicolumn{6}{|l|}{Overall score (out of 10)}	&	5.4	&	5.7	&	6.5	\\

\hline																	
\end{tabular}}
\label{table-calif3}
\end{table}
\section{Conclusions}
We introduced a new 3 sites  flexible model, the  FBA/$\epsilon$ which
 was parametrized using the experimental
values of the density, the dielectric constant and the dipole moment
at $1\;bar$ and $240\;K$.

This approach  gives a bond and angle distribution comparable
with experimental results. It is also  able to produce good agreement 
with the experiments  for thermodynamic and structural properties
at low and high temperatures. In particular the flexibility allow for 
the model to provide a good coexistence region of the density versus 
temperature phase diagram. The major advantage of the model is 
that is able to reproduce with accuracy a wide range of properties
what at different temperature.
This robust behavior makes it a good candidate for studying mixtures
of water and other polar materials.

\newpage
{\bf Acknowledgements}\\
 
 We thank the Brazilian agencies CNPq, INCT-FCx, and Capes for the
financial support. We also thank the SECITI of Mexico city for financial support.


\end{document}